\documentstyle[psfig,manuscript,aps]{revtex}
\tightenlines
\begin{document}

\title
{Screened Coulomb interactions in metallic alloys: II Screening beyond
the single-site and atomic sphere approximations}

\author{A.~V. Ruban}

\address{Center for Atomic-scale Materials Physics and Department of Physics,
\\ Technical University of Denmark, DK-2800 Lyngby, Denmark}

\author{S.~I. Simak}

\address{Department of Applied Physics, Chalmers University of Technology \\
and G{\"o}teborg University, S-41296, Gothenburg, Sweden }

\author{P.~A. Korzhavyi}

\address{Applied Materials Physics, Department of Materials Science and 
Engineering, \\
Royal Institute of Technology, 100 44 Stockholm, Sweden}

\author{H.~L. Skriver}

\address{Center for Atomic-scale Materials Physics and Department of Physics, 
\\ Technical University of Denmark, DK-2800 Lyngby, Denmark}

\date{2 August 2001}

\maketitle

\begin{abstract}
A quantitative description of the configurational part of the total energy
of metallic alloys with substantial atomic size difference cannot be 
achieved in the atomic sphere approximation: It needs to be corrected at 
least for the multipole moment interactions in the Madelung part of the 
one-electron potential and energy. In the case of a random alloy such
interactions can be accounted for only by lifting the atomic sphere
and single-site approximations, in order to include the polarization due 
to local environment effects. Nevertheless a simple parameterization of 
the screened Coulomb interactions for the ordinary single-site methods,
including the generalized perturbation method, is still possible. We 
obtained such a parameterization for bulk and surface NiPt alloys, which
allows one to obtain quantitatively accurate effective interactions in
this system.
\end{abstract}

\vspace{10mm}
\narrowtext

\section{Introduction}

One of the main problems of modern alloy theory is to establish a
{\it quantitatively} accurate description of the configuration-dependent part
of the free energy, i.e., the difference in the total energies of alloys with 
different atomic arrangements on the underlying lattice, in terms of 
effective cluster interactions which may subsequently be used in statistical 
thermodynamics simulations \cite{defontaine79,ducastelle91,zunger94}. Even 
without lattice relaxation effects (which are not considered here, although 
they play an important role in the phase equilibria of many alloy systems) 
a solution to the problem is still a challenge especially in the case of 
inhomogeneous systems such as surfaces in the presence of long-range and 
multi-site interactions which cannot be neglected.

The challenge originates from the fact that quantitatively accurate and 
reliable (within the accuracy of the approximation for the 
exchange-correlation part of the total energy of the electronic subsystem
in density functional theory (DFT) \cite{hohenberg64,kohn65}) effective 
cluster interactions can be obtained only by the Connolly-Williams (CW) or 
structure inversion method \cite{connolly83,zunger94} on the basis of the 
total energies of a set of specifically chosen ordered structures calculated 
by the so-called full potential (FP) methods, which have no restrictions on 
the form of the one-electron potential and density. However, the structure 
inversion methods become practically unusable in the case of an inhomogeneous
system, not only because of the large number of basic structures which must
be calculated to extract position- (layer-, for instance) dependent 
interactions, but mainly because of the large size of those basic structures 
(supercells) which are needed to factorize a specific interaction inside 
{\it homogeneous} parts of the systems, e.g., inside the layers parallel to 
a surface.

In this situation there appears to be only one alternative to the structure
inversion methods: The so-called generalized perturbation method (GPM) 
proposed by Ducastelle {\it et al.} \cite{ducastelle76,ducastelle91} on the 
basis of the coherent potential approximation (CPA) \cite{soven67,taylor67,%
kirkpatrick70} and formulated within tight-binding (TB) theory. Later the GPM
was generalized in a straightforward manner \cite{gonis87,turchi88,drchal92,%
singh93} for use in {\it ab initio} calculations based either on the 
Korringa-Kohn-Rostoker (KKR) method or the linearized muffin-tin orbitals 
(LMTO) method in the atomic sphere approximation (ASA). The main idea behind 
the GPM is to calculate perturbatively the total energy difference between 
the alloy in the initial state, which is completely random, and in a final 
state in which only one specifically chosen atomic distribution correlation 
function or short-range order parameter is different from that in the random 
state. This makes the GPM very efficient and convenient to use as it directly 
yields the needed effective cluster interactions.

However, it is known, although rarely mentioned in the literature, that the
interactions obtained by the GPM yield a quantitatively poor description of 
the ordering in real alloys (see, for instance, Ref. \cite{bose97}) in those
cases where there is a substantial size mismatch between the alloy components.
This failure may only partly be attributed to lattice relaxation effects. 
Rather, it originates not from the GPM method itself but is a consequence of 
inappropriate approximations in the basic methods underlying the GPM 
calculations. This is so, because, as has been demonstrated by Bieber {\it et 
al.} \cite{bieber83} in parametrized tight-binding calculations and by Singh 
{\it et al.} \cite{singh93x} in {\it ab initio} KKR-CPA calculations, the 
GPM interactions may provide (under certain conditions) a {\it consistent} 
description of the ordering or configurational energy. That is, the ordering
energy obtained from the GPM interactions, calculated in the framework of a
particular technique, agrees reasonably well with the ordering energy 
obtained directly from the total energy calculations by the {\it same} 
technique. 

The {\it ab initio} techniques underlying GPM calculations are usually the 
KKR-ASA, KKR-ASA-CPA, and LMTO-CPA methods \cite{gyorffy78} which are based 
on a number of approximations, such as the CPA, the single-site (SS) 
approximation for the electrostatic part of the DFT problem, and the 
spherical approximation for the form of the potential which, depending on 
the geometry, is called either the muffin-tin (MT) or atomic sphere 
approximation (ASA). The question is which of these approximations is the 
most severe in the cases where the alloy components have a substantial size 
difference. To answer the question, we note that a size difference leads to 
so-called "charge transfer effects" or, to be more precise, to a non-zero 
net charge for each alloy component inside their atomic spheres {\it chosen 
to be of equal size}.  

Although there are systems, where the CPA may lead to substantial errors, it
is clear from a general point of view that the CPA cannot be responsible for 
the errors in the case of pronounced charge transfer effects, because the 
error of the CPA is mainly related to specific features of the electronic 
structure of the individual alloy components such as the difference in the 
position and overlap of the energy bands \cite{ducastelle91,faulkner82}. 
Moreover, there is a number of different calculations which show that in such
systems as, for instance, CuPd and CuAu, where the alloy components have
similar electronic structures but different atomic size, the CPA works 
fairly well \cite{banhart89,abrikos98,kuqo98} as a method for obtaining 
the average electronic structure of random alloys. 

As far as the underlying KKR-CPA or LMTO-CPA methods are concerned, much 
larger errors may in fact come from the use of the single-site approximation 
in the self-consistent DFT part of the calculations as this yields no 
information about the distribution of the charge outside the individual 
atomic spheres of the alloy components. In fact, the effective medium outside
the individual atomic spheres of the alloy components is electroneutral, and 
therefore, if the net charge of an atomic sphere is non-zero, Poisson's 
equation cannot be solved properly. A number of different models have been 
proposed to include the missing {\it screening} charge in the solution 
to Poisson's equation \cite{gunnarsson83,koenig86,abrikos92,johnson93} and, 
most recently, a general formalism of screened Coulomb interactions (SCI) 
based on the knowledge of the spatial distribution of the screening charge 
around an impurity has been developed in Ref.\ \cite{ruban01} together with 
a formalism for the SCI contribution to the GPM interactions 
\cite{ducastelle91}. Although the SCI may now be included in SS-DFT-CPA 
calculations, this does not solve all problems connected with the description 
of the energetics of alloys.

It is not surprising that the main source of inaccuracy in the KKR-CPA and
LMTO-CPA methods is the spherical approximation, MT or ASA, for the form of 
the electron density and potential (in the following we will consider only 
the ASA, since the difference between the ASA and MT is unimportant for the 
later discussion and results). For instance, in the extreme case, where one 
of the alloy components is a vacancy, the error due to the use of the ASA is 
about 100\%  (or several eV in absolute values) for the vacancy formation 
energy \cite{braun91}. As has been shown by Korzhavyi {\it et al.} 
\cite{korzhavyi99}, this kind of error originates from the oversimplified 
description of the non-spherical electrostatic contribution to the 
one-electron potential and energy mainly from the charge density on the atoms 
{\it next} to the vacancy.

This is similar to the case of surfaces where the quite large ASA error may 
be substantially reduced by the inclusion of the multipole moments of the 
electron charges inside the atomic spheres \cite{skriver91}. The so-called 
ASA+M approach significantly improves vacancy and defect formation energies 
\cite{korzhavyi99,korzhavyi00}, surface energies \cite{ruban99} and alloy 
energetics \cite{smirnova01}. Recently, Finnis {\it et al.} \cite{finnis98} 
have included the multipole moments in their self-consistent tight-binding 
model which allowed them to obtain quite accurate description of the 
energetics of zirconia.

In this paper we show that the use of the ASA+M approach leads to a 
representation of the configurational part of the total energy, which is very
close to the full-potential results. Since the polarization of the atoms in 
an alloy is almost entirely determined by their closest local environment, it
is obvious that the effect of polarization cannot be described properly in 
the single-site approximation. Nevertheless, the SS-DFT-CPA methods may be 
still used for the electronic structure and total energy calculations of 
random alloys if the definition of the SCI is modified. It is the main 
purpose of the present paper to demonstrate how this may be done in the cases 
of ordinary bulk homogeneous random alloys and inhomogeneous systems such as
surfaces. 

The paper is organized as follows. In section II we introduce the ASA+M
approximation and outline some details of our calculations. In section III
we compare the ordering energies of NiPt alloys, calculated by the KKR
method in different approximations and by the Vienna {\it ab initio} 
simulation package (VASP) \cite{kresse93,kresse96}. In section IV we define 
the on-site screening Madelung potential, which should be added to the 
one-electron potential in the SS-DFT calculations. The SCI and the Madelung 
energy of a random alloy are defined in section V. In section
VI we calculate the intersite SCI in NiPt in different approximations
and by different methods. In section VII the screened generalized 
perturbation method interactions are calculated and compared with 
the Connolly-Williams interactions. In section VII we show how the formalism
for the SCI should be modified in the case of inhomogeneous systems where
there are several non-equivalent sublattices, like partially ordered alloys
or surfaces.

\section{Beyond the ASA}

In a previous publication \cite{ruban01}, in the following referred to as I,
we presented a consistent and variational, within DFT, approach to the 
electrostatic screening effects in random alloys, and within the ASA we found
that these screening effects were almost independent of alloy composition, 
lattice spacing, and crystal structure. However, in those cases where the
alloy components have a substantial size difference one cannot obtain a 
quantitatively correct description of the configurational part of the total 
energy of metallic alloys within the ASA. One must therefore go beyond the 
spherical approximation at least for the density. In the following we will 
show how this may be done. All other detail of our approach may be found in 
I.

\subsection{Multipole correction to the atomic sphere approximation:
ASA+M}

The idea behind the multipole correction is simply to include those 
contributions to the electrostatic multipole-moment expansion of the 
intercell or Madelung part of the one-electron potential and energy, which 
are neglected in the ASA. If the multipole moments of the electron charge, 
$q^L_{\bf R}$, inside an atomic sphere centered at {\bf R} are defined as  
 
\begin{equation} \label{Q_L}
q^L_{\bf R} = \frac{\sqrt{4\pi}}{2l+1} \int_{S_{\bf R}} 
\left( \frac{r}{S_{\bf R}} \right)^l n_{\bf R}({\bf r_R}) 
Y_L(\hat r_{\bf R}) d{\bf r_R} - Z_{\bf R} \delta_{0l} ,
\end{equation}
where $L$ is short-hand for the $(l,m)$ quantum numbers, $S_{\bf R}$ the 
radius of the atomic sphere, $n_{\bf R}$ the non-spherical charge density, 
and $Y_L$ a real harmonic, the Madelung contribution to the one electron 
potential is given by
 
\begin{equation} \label{eq:VMad}
v^L_{M{\bf R}} = \frac{1}{S} \sum_{{\bf R}',L'}^{} \
M^{L,L'}_{{\bf R},{\bf R}'} \ q^{L'}_{\bf R'} ,
\end{equation}
while the Madelung energy which now includes the multipole-multipole 
electrostatic interactions between different lattice sites may be written 

\begin{equation} \label{eq:EMad}
E_{M} = \frac{1}{2S} \sum_{{\bf R},L}^{} \ q^{L}_{\bf R} \
\sum_{{\bf R}',L'}^{} \ M^{L,L'}_{{\bf R},{\bf R}'} \
q^{L'}_{\bf R'} .
\end{equation}
In these expressions, $M^{L,L'}_{{\bf R},{\bf R}'}$ is the multipole Madelung 
matrix which is equivalent to the conventional LMTO structure constants and 
the number of multipoles included in the $L,L'$ summations is determined by 
the angular momentum cutoff $l_{max}$ in the basis set used in the Green's 
functions calculations. Owing to the properties of the Gaunt coefficients 
non-zero multipole moments of the charge density may be generated for 
$l$-values up to $l^M_{max}=2l_{max}$.

We note that, since in the ASA+M the one-electron potential is still kept 
spherically symmetric inside each atomic sphere, the only term which 
contributes to the one-electron potential is the L=(0,0) or monopole term. 
This simple restriction on the form of the-one electron potential violates 
the variational connection between the Madelung potential and energy and, in 
turn, between the one-electron potential and the total energy, i.e.,  

\begin{equation} \label{vE}
v_{M {\bf R}} \equiv  v^{00}_{M{\bf R}}
\neq \frac{\delta E_M}{\delta n_{\bf R}} .
\end{equation}
However, since this is just a consequence of the model, but not of theory in 
general, it does not create any problems. On the other hand, the 
reinstatement of the variational connection between the one-electron 
potential and the total energy by keeping only the monopole-multipole term in 
(\ref{eq:EMad}) may lead to a substantial error in the total energy, as 
the multipole-multipole interactions will not be accounted for.

\subsection{Details of calculations}

The Green's function technique has been used in both the KKR-ASA and the
locally self-consistent Green's function (LSGF) calculations in the scalar
relativistic and atomic sphere approximations. This part of the techniques 
is described in Refs.\ \cite{ruban01,ruban99,abrikosov97}. The basis 
functions have been expanded up to $l_{max}=3$ ($spdf$-basis) inside the 
atomic spheres, while the multipole moments have been calculated up to 
$l^M_{max}=6$. We have also performed a number of calculations in the 
$spd$-basis, i.e., $l_{max}=2$ and $l^M_{max}=4$. The integration of the 
Green's function over energy was performed in the complex plane over 16 
energy points on a semicircular contour using a Gaussian technique. We have 
used the generalized gradient approximation (GGA) of Perdew and co-workers 
\cite{perdew91}. For each structure the integration over the Brillouin zone 
(BZ) has been done by using equally spaced k-points in the irreducible part 
of the appropriate BZ and the number of k-point has been chosen to be 
equivalent to 500 - 1000 uniformly distributed k-points in the irreducible 
part of the BZ of the fcc structure. Core states were recalculated at 
each DFT iteration.

For benchmark calculations we applied the Vienna {\it ab initio\/} simulation
package (VASP) described in detail in Refs.\ \cite{kresse93,kresse96}. These 
calculations were performed in a plane-wave basis, utilizing fully non-local 
Vanderbilt-type ultrasoft pseudopotentials (US-PP) \cite{vanderbilt90} which 
allow the use of a moderate cutoff in the construction of the plane-wave 
basis for the transition metals. In the actual calculations the energy cutoff
was set to 302 eV, exchange and correlation were treated in the framework of 
the GGA \cite{perdew91}, and the integration over the Brillouin zone was 
performed on a Monkhorst-Pack k-mesh \cite{monkhorst72}. Test calculations 
showed that, depending on structure, the required convergence was reached for
35 to 275 k-points in the irreducible wedge of the BZ.

\section{Ordering energies in Ni-Pt}

We start by demonstrating the accuracy of the various approximations which 
are usually used in KKR(LMTO)-ASA-like calculations. For this purpose we have
calculated a set of ordered fcc NiPt alloys by the KKR-ASA method and by the 
US-PP for a fixed lattice constant without any local or anisotropic 
relaxations. The lattice constant has been chosen to be $a \approx$ 3.791 \AA, 
which corresponds to an atomic Wigner-Seitz radius of 2.8 a.u.. The ordered 
structures include: L1$_2$ (Cu$_3$Au-type), DO$_{22}$ (TiAl$_3$-type), Z3, 
$\beta$ \cite{lu91}, $\gamma$ (Pt$_2$Mo-type), L1$_0$ (CuAu-type), 
CH or "40" (NbP-type), Z2 \cite{lu91}, L1$_1$ (CuPt-type), and the so-called 
SQS-16 \cite{wolverton96}. 

To simplify the comparison we present in Table \ref{tbl:E_CWM} the values of 
the calculated "mixing" energies of the above mentioned ordered structures

\begin{equation} \label{eq:E_mix}
E_{mix}^{Ni_mPt_n} = 
E_{tot}^{Ni_mPt_n} - \frac{m E_{tot}^{Ni} + n E_{tot}^{Pt}}{m+n} ,
\end{equation}
where $E_{tot}^{Ni}$ and  $E_{tot}^{Pt}$ are the total energies of the pure 
components calculated at the same lattice constant. All total energies are
per atom. The mixing energies of the random alloys have been obtained on the 
basis of all the energies included in Table \ref{tbl:E_CWM}, except SQS-16, 
plus the energies of Ni$_7$Pt and Pt$_7$Ni (CuPt$_7$-type) ordered alloys 
(not presented in the Table) by the Connolly-Williams method in which the 
total energy expansion included pair interactions at the first four, 7th and 
10th coordination shells (these are the largest pair interactions in this as 
well as in many other fcc transition metal alloys), the first four triangle 
interactions and the two tetrahedron interactions corresponding to the 
tetrahedra of nearest neighbors, and the straight line along [111] direction
(the last being quite substantial in many systems).
 
The SQS-16 is a so-called special quasirandom structure \cite{zunger90} 
which consists of 8 atoms of one type and 8 atoms of another type distributed 
in the unit cell in such a way that the first 7 pair, the nearest neighbor 
triangle, and the tetrahedron atomic distribution correlation functions are
the same as in the random alloy. Hence, the fact that the values of 
$E_{mix}^{SQS-16}$ and $E_{mix}^{rand}$ are nearly eaqual indicates that: 1) 
the SQS-16 provides a good model for the random NiPt alloy, and 2) the 
convergence of the CW method is reasonably good. Part of the convergence of
the CW method is provided by the use of total energies on a fixed lattice, 
whereby the volume dependent contribution to the total energy is not expanded
in terms of cluster interactions, which is an ill-defined procedure in 
metallic systems and usually leads to very bad convergence of the
CW method (see, for instance, \cite{ozolins98}).

In Table \ref{tbl:E_CWM} we present KKR-ASA results in the $spdf$- as well
as the $spd$-basis and in both cases we show results in the pure ASA, i.e.,
without multipole moment contributions, and in the ASA+M. We also include the 
results of neutral sphere calculations, in which the atomic sphere of Pt is 
chosen to be larger than that of Ni in order to provide zero net charges of 
the atomic spheres. Although the comparison of the mixing energies should be 
done with some caution, because the ground state properties of the alloys 
are different in different approximations and because all the calculations 
have been performed at the same fixed lattice constant, it is clear that the 
ASA+M approach in the $spdf$-basis leads to values of $E_{mix}$ which are in
considerably better agreement with the US-PP results than any of the other 
approaches.

Using the results of the Connolly-Williams method for the total energy of
random alloys one can calculate the ordering energies, defined as the
difference between the total energies of an ordered and a random alloy at 
the same composition. In Table \ref{tbl:E_ord} we compare the ordering 
energies of different structures calculated in different approximations for 
Ni$_3$Pt, NiPt, and Pt$_3$Ni. Such a comparison makes sense since the 
ordering energies are much less volume-dependent than the total energies 
themselves. Again, it is seen that, relative to the US-PP results, the ASA+M 
approach in the $spdf$-basis gives not only the best, but also a quite 
accurate description of the ordering effects in NiPt. It is also seen that 
the ordering energies in the ASA+M but without multipole-multipole 
contribution to the Madelung energy appear to be half way beteeen ASA and
ASA+M results. As we will see below, this is in fact the limit of accuracy 
which can be reached in consistent single-site mean-field calculations.

It is also obvious from the table that the KKR-ASA does not yield reasonable 
values for the ordering energies in the case of transition metal alloys 
unless $f$-states are included in the basis. These $f$-states are needed to 
supply a better augmentation of the basis functions at the atomic sphere and 
a better interstitial charge density \cite{zhang92}. The neglect of 
$f$-states can only be partly compensated by the use of the so-called 
combine-correction term in the LMTO method \cite{andersen75}. 

Another important conclusion, which can be drawn from the results in Table 
\ref{tbl:E_ord}, is the fact that the use of neutral spheres leads to a 
substantial underestimate of the ordering effects in KKR-ASA(+M) 
calculations. In other words, although the neutral-sphere approach formally
solves the problem of the electrostatic interaction in an alloy, the 
electrostatic contribution to the one-electron potential and energy being 
zero by definition, it introduces errors which are unacceptable in a
quantitative description of the configurational energetics.

The reason for this failure is the following. If we compare the values of 
$E_{mix}$ from Table \ref{tbl:E_CWM} obtained with neutral spheres with 
those obtained with equal spheres, we find that the neutral sphere approach 
leads to substantial lowering of the total energy of the ordered alloys. 
However, the amount of lowering is structure dependent: In the case of 
"phase separated"-like structures, where there is a certain clustering of 
atoms of one type, like Z3, Z2, the lowering is much greater, than it is in
some of the more "normal" structures. Through the CW procedure the 
exaggerated lowering of the energies of the Z3 and Z2 structures leads to an
exaggerated lowering of the total energy of the random alloy, making, in the 
end, all the ordering energies much smaller in absolute value, than they
should be. In the case of the "phase-separated"-like structures this kind of 
error originates in the poor filling of the crystal space provided by 
the non-equal atomic spheres. The division of a crystal into atomic spheres 
is in itself quite a crude approximation, the error of which is not generally 
known. Now, if such division is done differently for different ordered 
structures by means of non-equal atomic spheres, it is obvious that the
resulting errors will be structure dependent and render a comparison of 
total energies meaningless.

Based on the above considerations it appears that the use of equal atomic 
spheres for the alloy components provides the only consistent and correct 
way of dividing space for configurational or ordering energy calculations,
if the underlying lattice is a simple Bravais lattice, such as fcc and bcc.
In that case the packing of space by the atomic spheres of the alloy 
components is homogeneous and independent of the alloy configuration
which makes at least part of the ASA error systematically compensated when 
the energy difference of different structures is calculated. 
Of course, the situation is different in alloys where the underlying lattice
has a more complicated structure. However, in the configurational energy
calculations one should follow the recipe of choosing equal sphere radii of 
the alloy components on the sublattice, where the alloying (ordering)
is taking place (see, for instance,\cite{korzhavyi01}).

\section{The on-site screening Madelung potential in the ASA+M}

Although the ASA+M approach seems a natural generalization of the ASA, that 
changes little in the formalism, it has a large effect on the way the SCI
must be treated in random alloys. The most dramatic consequence of the 
ASA+M is the fact that the correct SCI can no more be obtained in  
single-site electronic structure or Green's function calculations: The 
largest polarization effects, which give multipole-moment contributions to 
the SCI, actually comes from the nearest neighbors of the atom the screening 
of which is considered.

This is easy to see in the case of a single impurity in an otherwise perfect 
crystal which is the dilute limit of a random alloy. The point-group symmetry
of the impurity site is exactly the same as that of the underlying lattice,
while none of its neighbors has even inversion symmetry and, therefore, every 
atomic sphere around the impurity has a non-zero dipole moment. Such a dipole 
moment plus the higher multipole moments induced by the presence of the 
impurity can only be found in Green's function calculations which include
these neighboring sites in the perturbation part of the Dyson equation. 
Thus, the multipole-moment contribution to the SCI can be obtained only 
{\it beyond\/} the single-site approximation in the Green's function 
calculations. 

Nonetheless, we will determine a simple parameterization of the SCI for the 
SS-DFT-CPA method, which will allow us to obtain accurate results in 
SS-KKR(LMTO)-CPA calculations. Such a parameterization of the SCI in the 
ASA+M is purely a fitting procedure in contrast to the SCI determined by 
SS-ASA-Green's function calculations in I basically for the purpose of 
showing the existence of a consistent theory of the SCI in random alloys in 
the framework of the single-site-CPA theory.  

Let us first consider the on-site Madelung potential in a random alloy due to
the screened Coulomb interactions which should be used in SS-DFT-CPA
calculations for the random alloy. In the case of a binary random alloy, 
A$_{1-c}$B$_c$, this potential can be defined as (see I)

\begin{equation} \label{eq:V}
v_i = -e^2 \alpha_{scr}(0) \frac{q_i}{S} ,
\end{equation}
where $q_i$ is the net charge of the $i$th alloy component, $S$ the 
Wigner-Seitz radius, and $\alpha^i_{scr}(0)$ the on-site screening constant
which can be obtained from the screening charge by performing "impurity"
calculations (see I), for instance, by exchanging the type of atom at a 
particular site of the supercell modeling the random alloy. Then 
$\alpha_{scr}(0)$ is

\begin{equation} \label{eq:alpha_scr1}
\alpha_{scr}(0) = 
\frac{2S}{e^2}\sum_{L,\bf R} S_{0,L}({\bf R}) Q_L({\bf R}) ,
\end{equation}
where $S_{L,L'}({\bf R}')$ are the canonical structure constants, the first
few terms of which are explicitly defined, for instance, in Ref.\ 
\cite{andersen78}. Further, $Q_L({\bf R})$ are the normalized multipole 
moments of the screening charge in the atomic sphere centered at {\bf R}

\begin{equation}
Q_L({\bf R}) = \frac{\Delta q_L({\bf R})}{\Delta q({\bf R}=0)} 
\equiv \frac{\Delta q_L({\bf R})}{\Delta q_{L=00}({\bf R}=0)},
\end{equation} 
where $\Delta q_L({\bf R})$ is the difference between the L-moments of the
charge in the atomic sphere at ${\bf R})$ after and before the impurity has 
been introduced at the site ${\bf R}=$0.

There is one important point. In the pure ASA $\alpha_{scr}(0)$ can be 
determined in an alternative although formally equivalent manner by

\begin{equation} \label{eq:alpha_scr2}
\alpha_{scr}(0) = (\alpha_{rand}) = -\frac{S}{e^2} \frac{<v_i>}{<q_i>} ,
\end{equation}
where $<V_i>$ and $<q_i>$ are the average values of the Madelung potential
and net charges of the $i$th alloy component in the self-consistent
supercell calculations of the random alloy. Being practically exact in the 
ASA this scheme is only approximately valid in the ASA+M, where the average 
Madelung potential of the supercell, which is equal to the Madelung
potential of the underlying lattice, is not equal to zero in general due 
to the presence of the non-zero multipole moments at least for $l=$4. In
the case of inhomogeneous systems like surfaces, where the multipole 
contribution to the average Madelung potential is quite large, Eq.\  
(\ref{eq:alpha_scr2}) cannot be used at all. Therefore, (\ref{eq:alpha_scr2})
should be modified by substracting the corresponding average values of the 
net charges and Madelung potentials of the alloy components on the 
(sub)lattice, $\overline{q} = (1-c) <q_A> + c <q_B>$ and 
$\overline{v} = (1-c) <v_A >+ c <v_B>$

\begin{equation} \label{eq:alpha_scr3}
\alpha_{scr}(0) = (\alpha_{rand}) = -\frac{S}{e^2} 
\frac{<v_i>-\overline{v}}{<q_i>-\overline{q}}.
\end{equation} 

\begin{figure}
\centerline{\psfig{figure=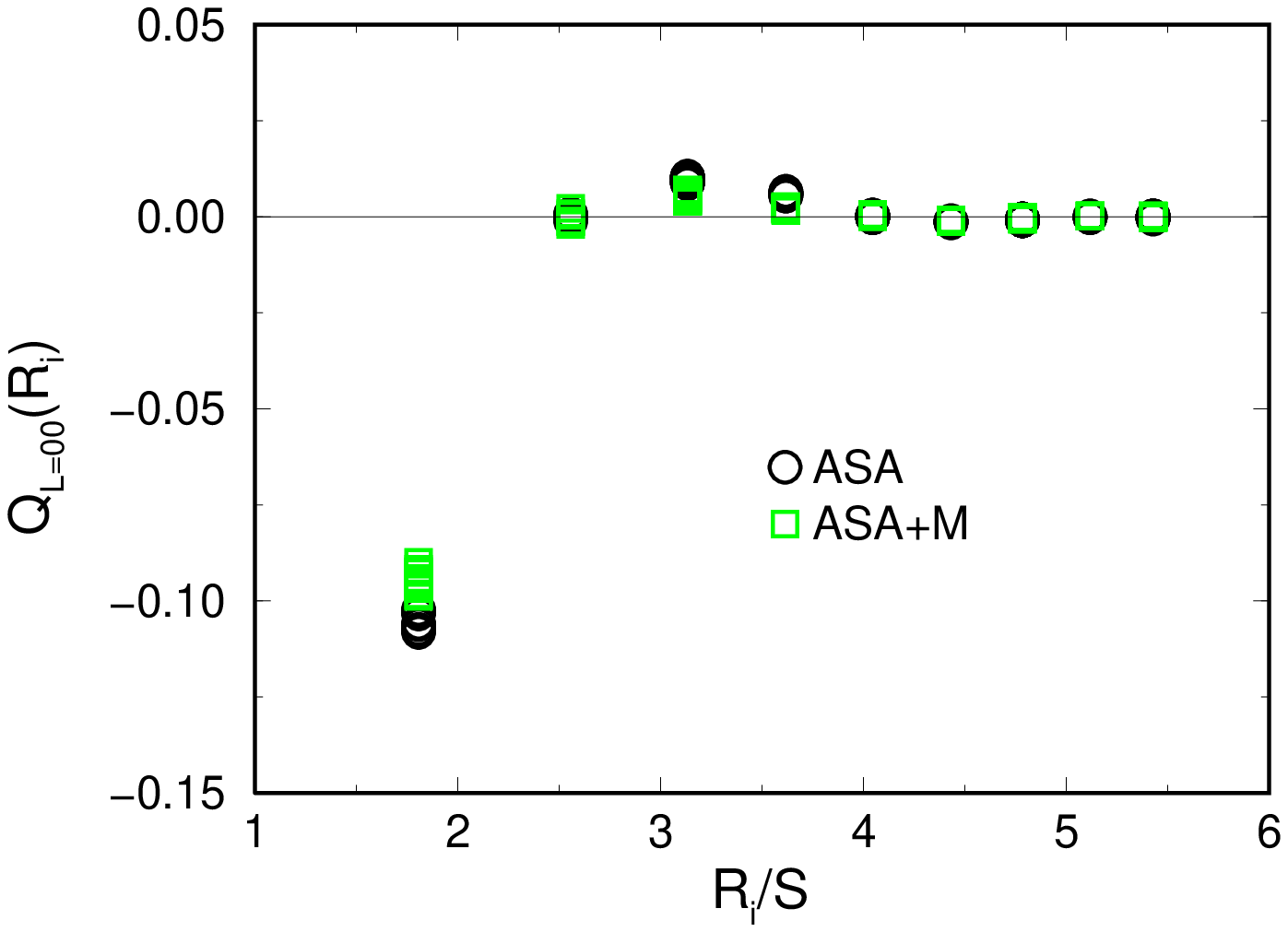,height=7.0cm}}
\caption[1]{The normalized screening charge in fcc NiPt, obtained 
in the ASA and ASA+M approximations. }
\label{fig:q_scr}
\end{figure}

In Fig.\ \ref{fig:q_scr} we compare the monopole moment of the normalized
screening charge, $Q_{L=00}(R)$, in a Ni$_{50}$Pt$_{50}$ random alloy 
obtained by changing the type of the atom from Pt to Ni at a some particular 
site in a 384-atom Ni$_{50}$Pt$_{50}$ supercell, the first seven short-range 
order (SRO) parameters of which are equal to zero. The calculations have been 
carried out in the single-site (SS)-LSGF-ASA as well as the embedded-cluster 
(EC)-LSGF-ASA+M methods (see I). The local interaction zone (LIZ) in the 
EC-LSGF-ASA+M calculations included 43 atoms, i.e., the central atom and its 
first three coordination shells (LIZ=4), while in the SS-LSGF calculations 
the LIZ included only one atom (LIZ=1).

It may be seen from Fig.\ \ref{fig:q_scr} that the screening is more 
efficient in the ASA+M than in the ASA, although the difference between the 
two cases is very small. Nevertheless, the effect of the multipole moments on
the on-site screening constant $\alpha_{scr}(0)$ is quite pronounced, it
increases to about 0.74 from the 0.61 in the ASA calculations. The largest
multipole contribution comes from the dipoles in the first coordination 
shell, which contribute almost 0.1 to the $\alpha_{scr}(0)$, while the 
quadrupole and octuple moments contribute 0.036 and 0.016, respectively.

The effective charge transfer, $\Delta q = <q_{Ni}> - <q_{Pt}>$, increases 
from 0.505 in the ASA to 0.583 in the ASA+M. However, it is still reproduced 
correctly in SS-DFT-CPA calculations, provided the {\it correct\/}, i.e.,
corresponding to the ASA+M, value of the on-site screening constant, 
$\alpha_{scr}(0)$, is used in (\ref{eq:V}). This is so, because the on-site 
screening constant is a parameter which determines a constant shift of the 
one-electron potential both in the ASA and ASA+M. This constant may be chosen
to contain the monopole-multipole interactions of the charge at a given site 
with its screening cloud, thereby yielding the correct effective charge 
transfer. Unfortunately, the same is not the case for the Madelung energy 
and the SCI in general.

\section{The SCI and Madelung energy of a random alloy}

The screened Coulomb interactions, $V_{scr}(R)$, are the energies of the
electrostatic interaction between the electron charge density inside an 
atomic sphere centered at some site (it is convenient to choose this site as
the origin) and the perturbed electron density and its screening charge at 
another, in general, different site. In the ASA+M, the SCI can be represented
in the form of the multipole expansion,

\begin{equation}
V_{scr}(R) = \sum^{}_{L} V^{L}_{scr}(R) \quad ,
\end{equation}
where $V^L_{scr}({\bf R})$ is the $L$-component contribution to the SCI at 
the distance $R$, which for a binary A$_{1-c}$B$_c$ alloy on a Bravais 
lattice may be expressed as in I

\begin{eqnarray} \label{V_scr^L}
V^L_{scr}({\bf R}) 
&=& \frac{e^2}{2} \Delta q^2 
\frac{Q_L({\bf R}=0)\alpha^L_{scr}({\bf R)}}{S} \\ \nonumber
&=& \frac{e^2}{4S}\Delta q^2 Q_L({\bf R}=0) 
\sum_{L',{\bf R}' \neq 0} S_{L=00,L'}({\bf R}') Q_{L'}({\bf R}'-{\bf R}) .  
\\ \nonumber 
\end{eqnarray}
Here $\Delta q$ is the difference of the net charges in the atomic sphere 
after and before the perturbation, i.e., the exchange of the type of atom at 
site ${\bf R}$, $\alpha^L_{scr}({\bf R})$ is a generalized screening constant,
$S_{L,L'}({\bf R}')$ are the canonical structure constants \cite{andersen78},
$Q_L({\bf R}'-{\bf R})$ are the normalized multipole moments of the screening 
charge, and $\Delta q_{L=00} = <q_A> - <q_B>$ is the effective charge 
transfer in the alloy.

Within the single-site mean-field considerations presented in I, all the 
multipole moments on the alloy sites are uncorrelated, the average value of 
$Q_L({\bf R}=0)$ being either very small or equal to zero, unless $L=00$,
and thus the only non-zero SCI is $V^{L=00}_{scr}$, which, for instance, 
in the case of the fcc underlying lattices can be written in the form (see I)

\begin{equation} \label{eq:V_scr(R)}
V_{scr}(R) = \Delta q^2 \frac{e^2}{2} \frac{\alpha_{scr}(R)}{S} ,
\end{equation}
where

\begin{equation} \label{eq:alpha}
\alpha_{scr}(R) = \frac{1}{2} \sum_{L',{\bf R}' \neq 0} 
S_{L=00,L'}({\bf R}') Q_{L'}({\bf R}'-{\bf R}) .  
\end{equation}

The on-site term $V_{scr}(R=0)$ is the energy of the electrostatic 
interaction between the net charge of an alloy component and its screening
density or, as has been shown in I, it is the screening Madelung energy of 
the random alloy. It is easy to see that this energy is DFT-consistent with 
the corresponding screening Madelung shift of the one-electron potential 
(\ref{eq:V}). However, in contrast to the screening Madelung potential which 
correctly reproduces the effective charge transfer in random alloys,
$V_{scr}(R=0)$ underestimates the corresponding Madelung energy in the ASA+M 
because of the missing contribution from the multipole-multipole interactions
in the single-site mean-field approximation. In the case of, for instance, 
a Ni$_{50}$Pt$_{50}$ random alloy this contribution is about $-$3 mRy/atom 
for the Wigner Seitz radius $S=$2.8 a.u..  

This means that if one wants a quantitatively accurate value of the total 
energy of a random alloy in the SS-DFT-CPA calculations consistent with the 
supercell ASA+M calculations one needs to modify the definition of the 
Madelung energy of the random alloy by introducing a fitting parameter. The 
simplest way to do so is to define the Madelung energy of the random alloy
as 

\begin{equation}
E_{Mad}^{rand} = \sum_{i} c_i E^{scr}_i,
\end{equation}
where $c_i$ is the concentration of the $i$th alloy component, and

\begin{equation}
E^{scr}_i = \frac{e^2}{2} \beta q_i^2 \frac{\alpha_{scr}(R=0)}{S} ,
\end{equation}
which means that in the case of a binary random alloy 
\begin{equation}
E_{Mad}^{rand} = c(1-c) \beta V_{scr}(R=0).
\end{equation}
Here, $\beta$ is the renormalization coefficient which is approximately equal
to 1.16 for most fcc and hcp transition metal random alloys. Thus, 
$E^{scr}_i$ and $V_{scr}^i$ are no more DFT-consistent (see (\ref{vE})).

As discussed above, this violation of general theory is a consequence of the 
ASA+M, which on the other hand brings the ordering energies of the much more
efficient SS-DFT-CPA approach into good quantitative agreement with the 
corresponding full-potential results. Although the difference between
the multipole-multipole and monopole-multipole results in Table 
\ref{tbl:E_ord} might not look so dramatic, the omission of the 
multipole-multipole interactions in the ASA+M Madelung energy has much 
more serious consequences in the case of, for instance, surface energy
anisotropy calculations, which cannot be reproduced even {\it qualitatively} 
without this term.

\begin{figure}
\centerline{\psfig{figure=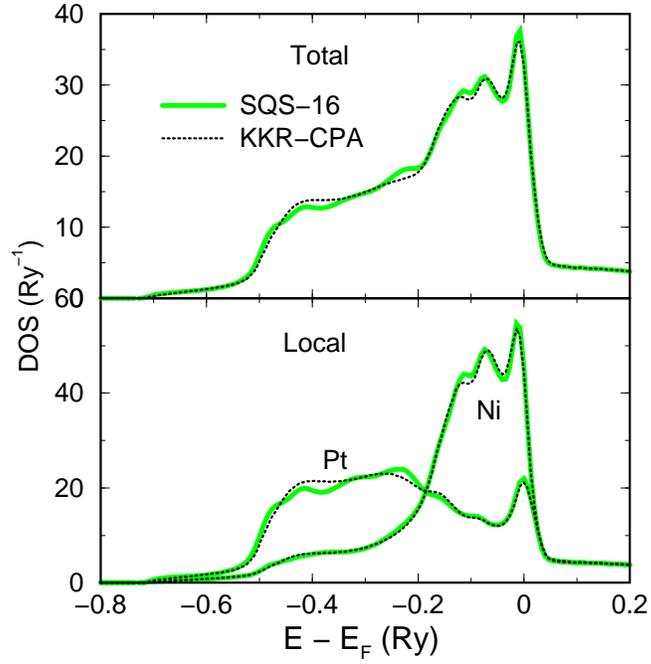,height=10.0cm}}
\caption[2]{The total and site-projected density of states in 
Ni$_{50}$Pt$_{50}$ obtained in the supercell EC-LSGF calculations
and by the SS-KKR-CPA method.}
\label{fig:DOS}
\end{figure}

Finally, we show that the CPA itself introduces relatively small errors in 
the electronic structure calculations of a random alloy. In Fig.\ 
\ref{fig:DOS} we compare the total density of states (DOS) of a random 
Ni$_{50}$Pt$_{50}$  alloy and the local, Pt and Ni, contributions obtained by 
two different methods, SS-KKR-CPA and EC-LSGF. The SS-KKR-CPA calculations 
have been performed with $\alpha_{scr}=0.74$ for the screening Madelung 
potentials of the alloy components (\ref{eq:V}), and in the EC-LSGF 
calculations LIZ=4 has been used to calculated the SQS-16 supercell of 
Ni$_{50}$Pt$_{50}$. Since, all the SRO parameter of the SQS-16 are equal to 
zero up to the 7th coordination shell, beyond which the net charge at each 
site is practically completely screened, the EC-LSGF results may be 
considered as a benchmark for the SS-KKR-CPA method. The agreement between 
the EC-LSGF and SS-KKR-CPA results is seen to be good clearly indicating 
that the CPA works fairly well for this system.

\section{Intersite screening constants in NiPt}

There are several ways of obtaining the intersite screening constants, 
$\alpha_{scr}(R)$, which determine the corresponding intersite SCI. 
First, they may be obtained directly from the normalized moments of the 
screening charge, $Q_L({\bf R})$, by menas of (\ref{eq:alpha}). This requires
two self-consistent supercell calculations: One performed for some initial 
atomic configuration in the supercell and the other for the same supercell 
with an "impurity" at the site where the type of atom is changed (see paper 
I). In this manner, however, only the monopole-multipole part of the SCI can 
be found.
 
Secondly, one may take advantage of the special properties of the LSGF 
method, namely of the fact that the electronic structure in the LSGF method 
is obtained in the so-called combined cluster-effective medium approach 
\cite{ruban01,abrikosov97}. That is, the local environment effects are taken
into account only inside the LIZ during the electronic structure calculations,
while the rest of the crystal is seen by each atom as a random alloy, 
described by the CPA effective medium. This means that all the correlated 
atomic configurations attributed to the non-zero SRO parameters beyond the 
LIZ do not contribute to the electronic structure of the supercell. At the
same time, since the Madelung problem is solved exactly, they are accounted 
for in the electrostatic part of the total energy.

Considering the change of the atomic configuration on the lattice as a small 
perturbation, one may argue on the basis of Andersen's force theorem 
\cite{mackintosh80} that the difference in the total energies of the alloy 
with some non-zero SRO parameters and a completely random alloy in the LSGF
calculations may be given by

\begin{eqnarray} \label{E_ord_lsgf}
E_{ord}^{LIZ=N}  &\equiv& E_{tot}^{SRO} - E_{tot}^{rand} \\ \nonumber 
&=& \frac{1}{2} c(1-c) \left[
\sum_{i=1}^{N-1} \alpha_i z_i V_{one-el}(R_i) + 
\sum_{i=1}^{N_{scr}} \alpha_i z_i V_{scr}(R_i) \right] ,
\end{eqnarray}
where $\alpha_i$ are the Warren-Cawley SRO parameters (see, for instance,
Ref.\ \cite{ducastelle91}) at the $i$th coordination shell, $N$ the size
of the local interaction zone, which is determined as the number of 
coordination shells around the impurity site plus 1, and $N_{scr}$ the 
coordination shell beyond which the SCI vanish. In the case of an fcc alloy
it may safely be assumed that, $N_{scr}=7$. It is obvious that the first term
in (\ref{E_ord_lsgf}) may be identified with the usual GPM interactions.

It follows from (\ref{E_ord_lsgf}), that if $N-1 < N_{scr}$ and 
$\alpha_i$=0 for all $i < N_{scr}$ except for one coordination shell, $j$, 
which is beyond the LIZ, $j > N-1 $, then the intersite SCI for this
specific coordination shell can be determined in two LSGF calculations:
One performed for a supercell which corresponds to the random alloy 
($\alpha_i = 0$ for all $i < N_{scr}$) and the other for a supercell
which satisfies the above described conditions with $\alpha_j \neq 0$. 
In this case

\begin{equation} \label{alpha_E_ord}
\alpha_{scr}(R_j) = \frac{4 S E_{ord}}{e^2 c(1-c) \Delta q^2 z_j \alpha_j} .
\end{equation}

In the ASA all the SCI can be determined by using (\ref{alpha_E_ord}) in
the corresponding SS-LSGF-CPA calculations, while $\alpha_{scr}(R_2)$ is 
the first SCI which can be determined in the ASA+M in this way, since 
there is no multipole contribution to the SCI in the single-site 
approximation (N(LIZ)=1). It is also clear that if the multipole-multipole
Madelung energy is included in the corresponding LSGF calculations, then,
in principle, $\alpha_{scr}(R)$ determined from (\ref{alpha_E_ord})
will also contain the multipole-multipole contribution. 

However, in this approach the higher order atomic distribution correlation 
functions (multisite SRO parameters) of the supercell should be also 
optimized. This is so, since all the multisite interactions for the figures 
inscribed in the LIZ with an vertex located at the central atom of the LIZ, 
also contribute to the ordering energy. Moreover, in the case of the ASA+M, 
in principle, there is a non-zero contribution from the SCI to the multisite 
interactions themselves. In our calculations, we have not optimized the
multisite SRO parameters of the supercells and, therefore, considering the
quite small values of the SCI beyond the first coordination shell we have 
not used (\ref{alpha_E_ord}) in the ASA+M calculations.

In Table \ref{tbl:SCI} we compare $\alpha_{scr}(R)$ for the first four 
coordination shells calculated either by direct summation of the normalized 
multipole moments (\ref{eq:alpha}) or by means of (\ref{alpha_E_ord}) in 
the SS-LSGF total energies. The LSGF calculations have been performed for 
384-atom supercells of an equiatomic Ni$_{50}$Pt$_{50}$ alloy with the 
corresponding sets of SRO parameters. Although we have not optimized the 
higher order SRO parameters, they turned out to be small: $\approx$ 0.02 or 
less, at least for the triangle and tetrahedra of the nearest neighbors on 
the fcc lattice. Thus they should not affect the results, at least in the 
ASA. As seen from Table \ref{tbl:SCI} the agreement between the intersite 
screening constants, $\alpha_{scr}(R)$, determined in two different 
calculations in the ASA, is quite reasonable.

As shown in I, $\alpha_{scr}(R)$ is practically a universal function for 
metallic alloys on simple Bravais lattices in the ASA and the single-site 
approximation. In the ASA+M approach, this is not the case any more. However,
our calculations for the dilute limit of NiPt alloys and for different 
lattice constants show that $\alpha_{scr}(R)$ changes very little in these 
cases and, in fact, less than the difference between the values of 
$\alpha_{scr}(R)$ calculated in two different ways. For instance, in
Table \ref{tbl:SCI} we show the values of $\alpha_{scr}(R)$ obtained
for a Ni impurity in pure Pt at $S=$ 3 a.u., which are very close to those
of the random equiatomic alloy at $S=$ 2.8 a.u.

\section{Screened GPM interactions in NiPt}

As demonstrated in I the intersite screened Coulomb interactions, 
$V_{scr}(R)$, given by (\ref{eq:V_scr(R)}) must be added to the corresponding
one-electron term given by the GPM in order to satisfy the force theorem 
\cite{mackintosh80}. In the case of a binary random alloy A$_c$B$_{1-c}$ 
\cite{multicomp} the screened generalized perturbation method (SGPM) 
interactions are defined as in Ref.\ \cite{ducastelle91}:
  
\begin{eqnarray} \label{SGPM}
V^{SGPM}(R) &=& V^{GPM}(R) + V_{scr}(R)  \\ \nonumber 
&=& V^{GPM}(R) + \frac{e^2}{2} \Delta q^2 \frac{\alpha_{scr}(R)}{S} ,
\end{eqnarray}
where $V^{GPM}(R)$ is the usual GPM interactions obtained as the change
of the one-electron energies due to specifically induced alloy configurations
on the alloy's underlying lattice. In fact, $V^{GPM}(R)$ should be 
renormalized due to the intersite SCI. However, this problem, as well as
the complete SGPM formalism will be considered elsewhere \cite{ruban}.

In Table \ref{tbl:V} we compare the first four most important effective pair 
interactions (the rest of the pair interactions and the multisite 
interactions are less than 0.1 mRy) in the NiPt fcc alloys at $S=$2.8 a.u. 
obtained by three different techniques, i) the SGPM for equiatomic alloy 
composition, ii) the Connolly-Williams method on the basis of the total 
energies of the ordered alloys described above, and iii) direct calculation
from the EC-LSGF total energies of Ni$_{50}$Pt$_{50}$ alloys similar to the
case of the intersite screening constant calculation described in the 
previous section. That is, in iii) the effective interactions have been 
obtained from the EC-LSGF total energy calculations for a completely random 
alloy, $\alpha_i=$0 for all $i < N_{scr}$, and for an alloy with one non-zero
$\alpha_j$. However, now $j < N-1$, where $N$ is the size of the LIZ in the 
LSGF calculations, and therefore the local environment effects attributed to
the non-zero SRO parameter, are included in the electronic structure 
calculations. In this case the effective pair interactions can be determined 
as

\begin{equation} \label{V_LSGF}
V_j = \frac{2 E_{ord}^{LIZ=N}}{c(1-c) z_j \alpha_j} .
\end{equation}
It is important, that (\ref{V_LSGF}) is not based on any additional 
approximations and therefore constitutes a direct way of determining the 
effective interactions with an accuracy which, in principle, is restricted 
only by the approximations used in the LSGF calculations, that is, mainly
by the ASA+M, since the CPA yields very small relative errors in the EC-LSGF 
calculations with LIZ$>2$. 
 
Although the SGPM interactions as well as the interactions determined from
(\ref{V_LSGF}) are concentration dependent, while the Connolly-Williams 
interactions are concentration independent, they can be compared since the 
concentration dependent interactions obtained for an equiatomic alloy 
composition are equal to those of the concentration independent interactions,
at the same fixed volume \cite{schweika89}. Therefore, if the basis in the 
Connolly-Williams method includes all the important interactions for a given
system, the Connolly-Williams interactions obtained from the KKR-ASA+M 
calculations of the ordered alloys at a fixed lattice constant should be 
equal to those obtained from (\ref{V_LSGF}) for the equiatomic alloy 
composition at the same lattice constant. 

It is clear from Table \ref{tbl:V}, that in general the agreement between the 
Connolly-Williams and the direct calculations is quite good. Further, the 
Connolly-Williams interactions obtained by the KKR-ASA+M and by the US-PP 
(CW-KKR-ASA+M and CW-US-PP) agree well with each other, except for $V_3$ 
which is a little larger in the CW-US-PP calculations than in the CW-KKR
calculations. This, then, confirms our point that the ASA+M approach allows 
us to obtain quantitatively accurate configurational energies of metallic 
alloys.

It is interesting to note that although the neutral-sphere approach yields
quite large errors for the ordering energies, it seems to work remarkably
well for the GPM interactions, except for $V_1$, which is more than twice 
as small as it should be. It is probably a coincidence that they come out 
very close to the CW-KKR-ASA+M interactions, since the SCI are quite small 
beyond the first coordination shell. Unfortunately, the CW interactions 
obtained in the neutral sphere calculations do not seem to be convergent,
the three- and four-site interactions being of the same order of magnitude 
as the interaction at the first coordination shell and therefore, they are 
not given in the table.

The SGPM interactions, obtained in the ASA+M and in the ASA (in these two 
cases we have used the intersite SCI calculated by (\ref{eq:V_scr(R)})) are
not very different, except for the interactions at the first coordination 
shell. This is most probably due to the missing multipole-multipole 
contribution to the intersite SCI in the ASA+M. One may also see from Table
Table \ref{tbl:V}, that the SGPM-KKR-ASA+M interactions are in fact quite 
close to those of the CW-KKR-ASA+M obtained without multipole-multipole 
electrostatic interactions (0-L). These interactions have recently been used 
in Monte Carlo simulations of the ordering in NiPt and reproduced quite well
the order-disorder transition temperature for an equiatomic alloy composition
and the values of the SRO parameters in a random alloy at T=1200 K 
\cite{pourovskii01}.

Finally, in Table \ref{tbl:E_ord_V} we show the ordering energies of the four 
equiatomic ordered alloys, obtained from the first 20 SGPM interactions, 
although the contribution from the interactions beyond the fourth 
coordination shell is only a few percent of the total ordering energy. 
Comparing these energies with those from the direct total energy 
calculations, presented in Table \ref{tbl:E_ord}, we find reasonable 
agreement for the ASA+M and the ASA results and very good agreement in 
neutral-sphere approach.

\section{The SCI at alloy surfaces}

The generalization of the SCI formalism to inhomogeneous systems, such as  
partially ordered alloys or surfaces, is straightforward. In the latter case
the SCI also become inhomogeneous and, therefore, definition 
(\ref{eq:V_scr(R)}) should be rewritten as

\begin{equation} \label{eq:V_scr(R)_tt}
V_{scr}^{\lambda \lambda'}(R) = \frac{e^2}{2} 
\Delta q_{\lambda} \Delta q_{\lambda'} 
\frac{\alpha_{scr}^{\lambda \lambda'}(R)}{S} ,
\end{equation}
where $\Delta q_{\lambda}$ is the effective charge transfer on the 
$\lambda$ sublattice, and $\alpha_{scr}^{\lambda \lambda'}(R)$ the 
screening constant, which is defined as

\begin{equation} \label{eq:alpha_tt}
\alpha_{scr}^{\lambda \lambda'}(R) = \frac{1}{2} \sum_{L',{\bf R}' \neq 0} 
S_{L=00,L'}({\bf R}') Q_{L'}^{\lambda}({\bf R}'-{\bf R}) . 
\end{equation}
Here, the vector ${\bf R}$ connects the site on sublattice $\lambda$, where
the perturbation of the charge density is induced to the site on sublattice
$\lambda'$, at which the SCI is determined (see paper I for details),
and $Q_L^{\lambda}({\bf R})$ are the multipole moments of the screening 
charge in the atomic sphere centered at ${\bf R}$ normalized by 
$\Delta q_{\lambda}(R=0)$.

It is clear that $\alpha_{scr}^{\lambda \lambda'}(R)$ depends on the
direction of ${\bf R}$, i.e., 
$\alpha_{scr}^{\lambda \lambda'}(R) \neq \alpha_{scr}^{\lambda' \lambda}(R)$,
and therefore the SGPM interactions, which in the inhomogeneous systems
should be invariant under a sublattice index interchange, are defined as 

\begin{eqnarray} \label{SGPM_surf}
V^{SGPM}_{\lambda \lambda'}(R) &=& V^{GPM}_{\lambda \lambda'}(R) +
\frac{1}{2} (V_{scr}^{\lambda \lambda'}(R) +  
V_{scr}^{\lambda' \lambda}(R)) = \\ \nonumber
&=& V^{GPM}_{\lambda \lambda'}(R) +
\frac{e^2}{4} \Delta q_{\lambda} \Delta q_{\lambda'} 
\frac{\alpha_{scr}^{\lambda \lambda'}(R) + 
\alpha_{scr}^{\lambda' \lambda}(R)}{S} , \nonumber
\end{eqnarray}
where $V^{GPM}_{\lambda \lambda'}(R)$ is the usual GPM interaction.

As an example of the inhomogeneous system we have chosen three low-index
fcc surfaces: (111), (100) and (110). The calculations has been performed by 
the EC-LSGF method with an inhomogeneous effective medium, which was fcc(111),
fcc(100), and fcc(110) slabs consisting of 3 vacuum and 6 atomic layers, 
4 vacuum and 6 atomic layers, and 6 vacuum and 10 atomic layers, respectively,
parallel to the surface. The actual supercells for the impurity calculations 
were built on the basis of the effective medium supercells by 
$N_x \times N_y$ translations in the plane parallel to the surface, which 
were $6 \times 6$ for the (111) and (100) surfaces, and $6 \times4$
for the (110) surface. In the case of the (110) surface $N_x$=6 was the 
period in the closed-packed [110] direction, while $N_y$=4 was the period in 
the [001] direction. Since the screening is insensitive to the alloy 
composition and the lattice parameter, the screening density was obtained for
a Ni impurity in pure Pt at $S=$3 a.u..

In Table \ref{tbl:alpha_0_surf} we present the on-site screening constant, 
$\alpha_{scr}^{\lambda}(0) \equiv \alpha_{scr}^{\lambda \lambda}(R=0)$,
in the first three layers of these surfaces (the impurity is in the 
$\lambda$ layer). It is clear that the surface makes the screening more 
efficient, since the larger value of the on-site screening constant means 
a closer position of the screening charge to the impurity, and, in fact, 
$\alpha_{scr}^{\lambda=1}(0)$ increases from the most close-packed (111) to 
the most open (110) surfaces. One can also see that the surface influences 
the screening mainly in the first layer for the fcc(111) and (100) surfaces, 
and very little in the second layer of the fcc(110) surface.

The intersite screening constants , $\alpha_{scr}^{\lambda \lambda'}(R)$, 
are presented in Table \ref{tbl:alpha_R_surf}. Again, one can see that 
the surface has a quite substantial effect on the screening constants 
$\alpha_{scr}^{1 \lambda'}(R)$, which is due to the perturbed electron
density in the first layer, $\lambda=$1. However, the values for 
$\alpha_{scr}^{\lambda \lambda'}(R)$ are already very close to the 
corresponding bulk values (see Table \ref{tbl:SCI}) for $\lambda=$2.

\section{Conclusion}

The polarization of the electron density of the alloy components due to their
size mismatch makes a substantial contribution to the electrostatic energy of 
the alloy. This contribution, which is missing in the pure ASA, may be 
accounted for in the ASA+M approach through the multipole-moment interactions 
in the Madelung part of the electrostatic problem, and, as we have shown, it 
plays a crucial role in obtaining the correct ordering energetics.

We have also demonstrated that the neutral sphere approach based on the
use of different atomic spheres for the alloy components on the corresponding
underlying lattice (or subllatice, in general) leads to unacceptable, 
quantitative errors, and therefore the only consistent way of obtaining
correct configurational energetics is to use spheres of equal radii for the 
alloy components on the sublattice where the alloying is taking place.

Since the multipole moments due to polarization effects originate from
the specific local atomic configuration around each site, they may in 
principle be accounted for only by methods which go beyond the single-site 
approximation in the electronic structure (Green's function) calculations.
However, the use of a simple parametrized form for the on-site Madelung 
potential and energy in the SS-DFT-CPA calculations still allows one to 
obtain a reasonably accurate description of the electronic structure (if the 
CPA works for a given system) and total energy, although, obviously, such a 
parameterization is possible only on the basis of the calculations by more 
accurate methods.

The monopole-multipole intersite SCI have been obtained for NiPt fcc bulk and
surface alloys. The SGPM interactions, which are the usual GPM interactions 
plus the SCI, reproduce the corresponding monopole-multipole KKR-ASA+M 
results, which give a semiquantitatively correct description of the ordering 
in NiPt.

\section{Acknowledgements}

Valuable discussions with Dr.\ I.~A.\ Abrikosov, are greatly acknowledged.  
Center for Atomic-scale Materials Physics is sponsored by the Danish National
Research Foundation. S.I.S. acknowledges support from the Swedish National
Research Council (NFR) and Materials Consortium "ATOMICS". The work of P.K.
is financed by the SKB AB, The Swedish Nuclear Fuel and Waste Management 
Company.

\noindent
\begin{table}
\caption{"Mixing" energies, $-E_{mix}$ (in mRy/atom) of ordered and random 
NiPt alloys obtained by different methods at a fixed Wigner-Seitz radius 
of S=2.8 a.u.}
\begin{tabular}{l|lllllc}
      &  \multicolumn{5}{c}{KKR-ASA}  &  \\
Alloy &  \multicolumn{3}{c}{$spdf$} & \multicolumn{2}{c} {$spd$} & US-PP \\
                   & ASA+M  & ASA   & neutral  & ASA+M &  ASA  &      \\
\hline
\hline
             & \multicolumn{6}{c} {Ni$_{25}$Pt$_{75}$} \\               
L1$_2$       & 21.7 & 19.5 & 28.4 & 31.1 & 31.7 & 22.3 \\
DO$_{22}$    & 20.6 & 18.3 & 26.4 & 30.3 & 30.1 & 21.0 \\
Z3           & 13.2 &  8.6 & 23.4 & 18.1 & 17.0 & 14.5 \\
random       & 15.6 & 11.7 & 23.7 & 22.0 & 21.1 & 16.4 \\
             & \multicolumn{6}{c} {Ni$_{33}$Pt$_{66}$} \\
"Pt$_2$Mo"   & 21.5 & 18.2 & 28.3 & 31.9 & 31.2 & 22.5 \\
$\beta$      & 16.4 & 10.7 & 27.7 & 23.7 & 22.5 & 19.4 \\
             & \multicolumn{6}{c} {Ni$_{50}$Pt$_{50}$} \\
L1$_0$       & 27.5 & 23.8 & 33.1 & 39.0 & 38.4 & 28.0 \\
Z2           & 13.4 &  5.1 & 25.1 & 17.0 & 13.4 & 15.0 \\
"CH"         & 26.3 & 22.8 & 31.0 & 38.0 & 37.4 & 26.9 \\
L1$_1$       & 22.7 & 17.5 & 32.8 & 30.8 & 30.1 & 23.0 \\
SQS-16       & 20.7 & $-$  &  $-$ & $-$  & $-$  & 21.7 \\
random       & 20.6 & 14.9 & 29.3 & 28.7 & 26.8 & 21.6 \\
             & \multicolumn{6}{c} {Ni$_{66}$Pt$_{33}$} \\
$\beta$      & 17.6 & 10.0 & 23.9 & 23.9 & 20.6 & 19.3 \\
"Pt$_2$Mo"   & 21.9 & 18.1 & 26.2 & 31.6 & 30.2 & 22.9 \\
             & \multicolumn{6}{c} {Ni$_{25}$Pt$_{75}$} \\ 
Z3           & 14.0 & 8.4  & 18.4 & 18.5 & 15.8 & 14.6 \\
DO$_{22}$    & 19.7 & 16.7 & 21.8 & 28.3 & 27.0 & 20.2 \\
L1$_2$       & 20.0 & 16.8 & 22.0 & 28.2 & 27.2 & 20.6 \\
random       & 15.4 & 10.7 & 20.4 & 21.1 & 19.1 & 16.4 \\
\end{tabular}
\label{tbl:E_CWM}
\end{table}

\noindent
\begin{table}
\caption{Ordering energies (in mRy/atom) NiPt alloys at a fixed Wigner-Seitz 
radius of S=2.8 a.u.. Ordering energies obtained without multipole-multipole
contribution in the ASA+M calculations are given in parentheses.}
\begin{tabular}{l|rrrrrr}
      &  \multicolumn{5}{c}{KKR-ASA}  &  \\
Alloy &  \multicolumn{3}{c}{$spdf$} & \multicolumn{2}{c} {$spd$} & US-PP \\
                   & ASA+M  & ASA   & neut.  & ASA+M &  ASA  &      \\
\hline
\hline
      & \multicolumn{6}{c} {Pt$_3$Ni} \\
L1$_2$       & -6.13 (-7.07) & -7.84 & -4.69 & -9.17 & -10.58 & -5.98 \\
DO$_{22}$    & -5.06 (-5.91) & -6.63 & -2.67 & -8.34 &  -9.04 & -4.63 \\
Z3           &  2.34 (2.72) &  3.06 &  0.31 & -3.85 &   4.09 &  1.83 \\
      & \multicolumn{6}{c} {NiPt} \\
L1$_0$       & -6.91 (-8.05) & -8.92 & -3.82 & -10.31 & -11.63 &  -6.49 \\
Z2           &  7.18 ( 8.66) &  9.85 &  4.14 &  11.72 &  13.35 &   6.55 \\
"CH"         & -5.66 (-6.86) & -7.87 & -1.67 & -10.05 & -10.68 &  -5.30 \\
L1$_1$       & -2.03 (-2.38) & -2.58 & -3.54 &  -2.15 &  -3.39 &  -1.47 \\
SQS-16       & -0.07 (-0.12) &   -   &   -   &    -   &    -   &  -0.18 \\
      & \multicolumn{6}{c} {Ni$_3$Pt} \\
L1$_2$        & -4.58 (-5.48) & -6.12  & -1.66 & -7.05 & -8.05 & -4.52 \\
DO$_{22}$     & -4.29 (-5.25) & -6.01  & -1.37 & -7.12 & -7.09 & -4.20 \\
Z3            &  1.35 (1.92) &  2.31  &  1.99 &  2.65 &  3.33 &  1.48 \\
\end{tabular}
\label{tbl:E_ord}
\end{table}

\noindent
\begin{table}
\caption{$\alpha_{scr}(R_i)$ at the first four coordination shells in the
ASA and ASA+M obtained by different methods.}
\begin{tabular}{l|cccc}
Method & i=1 &  2 &  3 &  4  \\ 
\hline
       & \multicolumn{4}{c} {ASA}          \\  
Eq. (13)                  &  0.1584 & -0.0017 & -0.0163 & -0.0108 \\
Eq. (18)                  &  0.1640 & -0.0026 & -0.0189 & -0.0116 \\
       & \multicolumn{4}{c} {ASA+M}            \\  
Eq. (13)                  &  0.1279 & -0.0023 & -0.0101 & -0.0050 \\
Eq. (13) (impurity)       &  0.1304 & -0.0035 & -0.0106 & -0.0052 \\
\end{tabular}
\label{tbl:SCI}
\end{table}

\noindent
\begin{table}
\caption{Effective interatomic interactions at the first four coordination
shells obtained by different methods (in mRy). The SCI contribution to the
SPGM interactions is given in parentheses.}
\begin{tabular}{l|llll}
Approximation & $V_1$ &  $V_2$ &  $V_3$ &  $V_4$  \\ 
\hline
       & \multicolumn{4}{l} {SGPM} \\
ASA+M (0-L)  & 14.05 (15.44) & 0.32 (-0.10) & -1.09 (-1.22) & -1.76 (-0.84) \\
ASA          & 12.26 (14.35) & 0.53 (-0.15) & -1.31 (-1.48) & -2.14 (-0.98) \\
Neutr. (GPM) &  5.49         & 1.22         &  0.01         & -0.73         \\
       & \multicolumn{4}{l} {Connolly-Williams method} \\
ASA+M        & 12.68         & 1.31         & -0.02         & -0.73         \\
ASA+M (0-L)  & 13.70         & 0.49         & -0.86         & -1.39         \\
ASA          & 14.33         & 0.28         & -1.72         & -1.92         \\

US-PP        & 12.81         & 1.30         &  0.69         & -0.40         \\
       & \multicolumn{4}{l} {Direct calculations from (19)}     \\
ASA+M        & 12.45         & 0.47         & -0.49         & -0.65         \\
\end{tabular}
\label{tbl:V}
\end{table}

\noindent
\begin{table}
\caption{Ordering energies of the equiatomic NiPt alloys obtained from
the pair SGPM effective interactions. The values obtained in the direct
calculations (see Table II) are given in parentheses.}
\begin{tabular}{l|rrr}
Structure &   ASA+M  &    ASA &  Neutr. \\
\hline
L1$_0$    &  -7.73 (-8.05) &  -7.46 (-8.92) &  -3.61 (-3.82) \\     
Z2        &   9.22 ( 8.66) &   8.68 ( 9.85) &   4.19 ( 4.14) \\  
"CH"      &  -7.15 (-6.86) &  -6.62 (-7.87) &  -1.77 (-1.67) \\ 
L1$_1$    &  -2.96 (-2.38) &  -3.75 (-2.58) &  -3.09 (-3.54) \\
\end{tabular}
\label{tbl:E_ord_V}
\end{table}

\noindent
\begin{table}
\caption{The on-site screening constant, $\alpha_{scr}^{\lambda }(0)$
in the first 3 layers of (111), (100) and (110) fcc surfaces.}
\begin{tabular}{l|ccc}
 facet & $\lambda$ =1 &  2 &  3   \\ 
\hline
 fcc(111)   &  0.805 &  0.730 &  0.728 \\
 fcc(100)   &  0.841 &  0.732 &  0.729 \\
 fcc(110)   &  0.840 &  0.756 &  0.732 \\ 
\end{tabular}
\label{tbl:alpha_0_surf}
\end{table}

\noindent
\begin{table}
\caption{$\alpha_{scr}^{\lambda \lambda'}(R_i)$ at the first four 
coordination shells of the (111), (100) and (110) fcc surfaces.}
\begin{tabular}{llrrrr}
$\lambda$ & facet & $\lambda'$ =1 &  2 &  3 &  4  \\ 
\hline
       \multicolumn{6}{c} {$R_1$}             \\  
1  &   fcc(111)   &  0.1160 &  0.1262 &   -     &    -   \\
   &   fcc(100)   &  0.1007 &  0.1241 &   -     &    -   \\
   &   fcc(110)   &  0.1026 &  0.1176 &  0.1285 &    -   \\ 
\hline
2  &   fcc(111)   &  0.1297 &  0.1295 &  0.1287 &    -   \\
   &   fcc(100)   &  0.1263 &  0.1322 &  0.1298 &    -   \\
   &   fcc(110)   &    -    &  0.1325 &  0.1306 &  0.1318 \\
\hline
3  &   fcc(111)   &    -    &  0.1319 &  0.1317 &  0.1307 \\
   &   fcc(100)   &    -    &  0.1325 &  0.1306 &  0.1318 \\
   &   fcc(110)   &  0.1290 &  0.1297 &  0.1285 &  0.1312  \\
\hline
\hline
       \multicolumn{6}{c} {$R_2$}             \\  
1  &   fcc(111)   &    -    &  0.0012 &   -     &    -    \\
   &   fcc(100)   & -0.0111 &    -    & -0.0022 &    -    \\
   &   fcc(110)   & -0.0111 &    -    &  0.0040 &    -    \\
\hline 
2  &   fcc(111)   & -0.0035 &    -    &  0.0012 &    -    \\
   &   fcc(100)   &    -    & -0.0003 &   -     &  0.0017 \\
   &   fcc(110)   &    -    & -0.0021 &   -     & -0.0011 \\
\hline
3  &   fcc(111)   &    -    & -0.0003 &   -     &  0.0001 \\
   &   fcc(100)   &  0.0027 &    -    &  0.0016 &   -     \\
   &   fcc(110)   & -0.0039 &    -    & -0.0023 &   -     \\
\hline
\hline
       \multicolumn{6}{c} {$R_3$}             \\   
1  &   fcc(111)   & -0.0139 & -0.0099 & -0.01441 &    -    \\
   &   fcc(100)   &    -    & -0.0080 & -0.0156  &    -    \\
   &   fcc(110)   & -0.0131 & -0.0090 & -0.0101  & -0.0111 \\
\hline
2  &   fcc(111)   & -0.0104 & -0.0096 & -0.0095  & -0.0111 \\
   &   fcc(100)   & -0.0097 &    -    & -0.0088  & -0.0096 \\
   &   fcc(110)   & -0.0079 & -0.0069 & -0.0107  & -0.0117 \\
\hline 
3  &   fcc(111)   & -0.0099 & -0.0097 & -0.0111  & -0.0113 \\
   &   fcc(100)   & -0.0091 & -0.0087 &    -     & -0.0103 \\
   &   fcc(110)   & -0.0124 & -0.0107 & -0.0118  & -0.0108 \\
\hline 
\hline
       \multicolumn{6}{c} {$R_4$}             \\ 
1  &   fcc(111)   & -0.0058 &    -    & -0.0087 &    -    \\
   &   fcc(100)   & -0.0043 &    -    & -0.0103 &    -    \\
   &   fcc(110)   & -0.0031 &    -    & -0.0063 &    -    \\ 
\hline
2  &   fcc(111)   &    -    & -0.0031 &   -     & -0.0042 \\
   &   fcc(100)   &    -    & -0.0033 &   -     & -0.0030 \\
   &   fcc(110)   &    -    & -0.0047 &   -     & -0.0036 \\ 
\hline
3  &   fcc(111)   & -0.0024 &    -    & -0.0034 &    -    \\
   &   fcc(100)   & -0.0019 &    -    & -0.0030 &    -    \\
   &   fcc(110)   & -0.0029 &    -    & -0.0058 &    -    \\ 
\end{tabular}
\label{tbl:alpha_R_surf}
\end{table}


\newpage
\begin{references}


\bibitem{defontaine79} D. de Fontaine, in {\it Solid State Physics},
edited by H. Erenreich, F. Seitz, and D. Turnbull (Academic,
New York, 1979), vol. {\bf 2}, p. 117.

\bibitem{ducastelle91} F. Ducastelle, {\it Order and Phase Stability in
Alloys} (North-Holland, Amsterdam, 1991).

\bibitem{zunger94} A. Zunger, in {\it NATO ASI on Statics and Dynamics of
Alloy Phase Transformations} (Plenum Press, New York, 1994), p. 361.

\bibitem{hohenberg64} P. Hohenberg and W. Kohn, Phys.\ Rev.\ 136B, 864 (1964).

\bibitem{kohn65} W. Kohn and L.J. Sham, Phys.\ Rev.\ 140A, 1133 (1965).

\bibitem{connolly83} J.  W.  D.  Connolly, and A.  R.  Williams, Phys.  Rev.
B {\bf 27}, 5169 (1983).

\bibitem{ducastelle76} F. Ducastelle and F. Gautier, J. Phys. F {\bf 6},
2039 (1976).

\bibitem{soven67} P. Soven, Phys. Rev. {\bf 156}, 809 (1967).

\bibitem{taylor67} D.~W. Taylor, Phys. Rev.  {\bf 156}, 1017 (1967).

\bibitem{kirkpatrick70} S. Kirkpatrick, B. Velicky, H. Erenreich,
Phys. Rev. B {\bf 1}, 3250 (1970).

\bibitem{gonis87} A. Gonis, X.-G. Zhang, A. J. Freeman, P. Turchi,
G. M. Stocks, and D. M. Nicholson, Phys. Rev. B {\bf 36}, 4630 (1987).

\bibitem{turchi88} P. E. A. Turchi, G. M. Stocks, W. H. Butler,
D. M. Nicholson, and A. Gonis, Phys. Rev. B {\bf 37}, 5982 (1988).

\bibitem{drchal92} V. Drchal, J. Kudrnovsk\'y, L. Udvardi, P. Weinberger,
A. Pasturel, Phys. Rev. B {\bf 45}, 14328 (1992).

\bibitem{singh93} P. P. Singh and A. Gonis, Phys. Rev. B {\bf 47}, 6744 
(1993).

\bibitem{bose97} S. K. Bose, V. Drchal, J. Kudrnovsky, O. Jepsen, 
O. K. Andersen, Phys. Rev. B {\bf 55}, 8184 (1997).

\bibitem{bieber83} A. Bieber, F. Ducastelle, F. Gautier, G. Treglia,
and P. Turchi, Soild State Comm., {\bf 45} 585 (1983) .

\bibitem{singh93x} P.~P. Singh, A. Gonis, E.~A. Turchi, Phys. Rev. Lett.,
{\bf 71} 1605 (1993).

\bibitem{gyorffy78} B.~L. Gyorffy and G.~M. Stocks in
{\it Electrons in Disordered Metals and at Metallic Surfaces} ed by
P. Phariseau, B.~L. Gyorffy, and L. Scheire, (New York: Plenum, 1978).

\bibitem{faulkner82} J.~S. Faulkner,
Prog.\ Mater.\ Sci.\  {\bf 27}, 1 (1982).

\bibitem{banhart89} J. Banhart, P. Weinberger, J. Voitl\"ander, 
Phys. Rev. B {\bf 40}, 12079 (1989).

\bibitem{abrikos98} I. A. Abrikosov and B. Johansson, Phys. Rev. B,
{\bf 57}, 14164 (1998).

\bibitem{kuqo98} M. Borici-Kuqo, R. Monnier, and V. Drchal,
Phys. Rev. B {\bf 58}, 8355 (1998).

\bibitem{gunnarsson83} O. Gunnarsson, O. Jepsen, and O. K. Andersen, 
Phys.\ Rev.\ B {\bf 27} 7144 (1983).

\bibitem{koenig86} C. Koenig, N. Stefanou,  and J. M. Koch, Phys. Rev. 
B {\bf 33}, 5307 (1986)

\bibitem{abrikos92} I.~A.  Abrikosov, Yu.~H.  Vekilov, P.~A.
Korzhavyi, A.~V. Ruban, and L.~E. Shilkrot,
Solid State Commun.\  {\bf 83} 867 (1992).

\bibitem{johnson93} D.~D. Johnson and F.~J. Pinski,
Phys. Rev. B {\bf 48} 11553 (1993).

\bibitem{ruban01} A. V. Ruban, H. L. Skriver, Phys. Rev. B, submitted

\bibitem{braun91} P. Braun, M. Fahnle, M. van Schilfgaarde, and
O. Jepsen, Phys. Rev. B, {\bf 44}, 845 (1991).

\bibitem{korzhavyi99} P. A. Korzhvyi, I. A. Abrikosov, B. Johansson,
A. V. Ruban, and H. L. Skriver, Phys. Rev. B {\bf 59} 11693 (1999).

\bibitem{skriver91} H. L. Skriver and N. M. Rosengaard,
Phys.\ Rev.\ B {\bf 43} 9538 (1991).

\bibitem{korzhavyi00} P. A. Korzhavyi, A. V. Ruban, A. Y. Lozovoi,
Yu. Kh. Vekilov, I. A. Abrikosov, B. Johansson, Phys. Rev. B {\bf 61} 
6003 (2000).

\bibitem{ruban99} A.~V. Ruban and H.~L. Skriver,
Comp. Mat. Sci. {\bf 15}, 119 (1999).

\bibitem{smirnova01} E. A. Smirnova, Yu. Kh. Vekilov, B. Johansson,
and I. A. Abrikosov, Phys. Rev. B {\bf 64}, 20102 (2001).

\bibitem{finnis98} M. W. Finnis, A. T. Paxton, M. Methessel, 
M. van Schilfgaarde, Phys. Rev. Lett. {\bf 81}, 5149 (1998).

\bibitem{kresse93}
G. Kresse and J. Hafner, Phys. Rev. B {\bf 48}, 13 115 (1993).

\bibitem{kresse96}
G. Kresse and J. Furthm{\"u}ller, Comp. Mater. Sci. {\bf 6},
15 (1996); Phys. Rev. B {\bf 54}, 11 169 (1996).

\bibitem{abrikosov97} I.  A.  Abrikosov, S.  I.  Simak, B.  Johansson, A.
V. Ruban, and H. L. Skriver, Phys. Rev. B {\bf 56}, 9319 (1997).

\bibitem{perdew91}
Y. Wang and J.P. Perdew, Phys. Rev. B {\bf 44}, 13 298 (1991);
J.P. Perdew, J.A. Chevary, S.H. Vosko, K.A. Jackson, M.R. Pederson,
D.J. Singh, and C. Fiolhais, Phys. Rev. B {\bf 46}, 6671 (1992).

\bibitem{vanderbilt90} D. Vanderbilt. Phys. Rev. B {\bf 41}, 7892 (1990).

\bibitem{monkhorst72}
H.J. Monkhorst and J.D. Pack, Phys. Rev. B {\bf 13}, 5188 (1972).

\bibitem{lu91} Z. W. Lu, S.-H. Wei, A. Zunger, S. Frota-Pessoa, and
L. G. Ferreira, Phys. Rev. B {\bf 44}, 512 (1991).

\bibitem{wolverton96} C. Wolverton, A. Zunger, S. Froyen, and S.-H. Wei,
Phys. Rev. B {\bf 54}, 7843 (1996).

\bibitem{ozolins98} V. Ozolins, C. Wolverton, and A. Zunger, Phys. Rev.
B {\bf 57}, 6427 (1998).

\bibitem{zunger90} A. Zunger, S.-H. Wei, L.~G. Ferreira, and J.~E. Bernard,
Phys. Rev. Lett. {\bf 65}, 353 (1990).

\bibitem{zhang92} X.-G. Chang, W. H. Butler, Phys. Rev. B {\bf 46},
7433 (1992).

\bibitem{andersen75} O. K. Andersen, Phys.\ Rev.\ B {\bf 12} 3060 (1975).

\bibitem{korzhavyi01} P.~A. Korzhavyi, L.~V.~Pourovskii,
H.~W.~Hugosson, A.~V.~Ruban, and B.~Johansson, to be published.

\bibitem{andersen78} O. K. Andersen, W. Klose, and H. Nohl, Phys. Rev.
B {\bf 17}, 1209 (1978).

\bibitem{mackintosh80} A. R. Mackintosh and O. K. Andersen, in
Electrons at the Fermi Surface ed by M. Springford, Cambridge: 1980.

\bibitem{multicomp} The generalization to the multicomponent systems is 
obvious, since the so-called generalized effective pair interactions, which
are given by the pair potentials of three different alloy components, are
expressed in terms of usual effective interactions, similar to that defined
in the case of a binary alloy - see appendix in A.~V. Ruban, H.~L. Skriver,
Phys. Rev. B {\bf 55}, 856 (1997).

\bibitem{ruban} A. V. Ruban, unpublished.

\bibitem{schweika89} W. Schweika, A. E. Carlsson, Phys. Rev. B {\bf 40},
4990 (1989).

\bibitem{pourovskii01} L. V. Pourovskii, A. V. Ruban, I. A. Abrikosov,
Y. Kh. Vekilov, and B. Johansson, Phys. Rev. B {\bf 64}, 35421 (2001).

\end{references}
\end{document}